\documentclass[preprint,12pt,authoryear]{elsarticle}

\usepackage[T1]{fontenc}
\usepackage{lmodern}
\usepackage{orcidlink}

\usepackage{amsmath}
\usepackage{amssymb}
\usepackage{amsthm}

\usepackage{graphicx}
\usepackage{epstopdf}
\usepackage[caption=false]{subfig}
\usepackage{booktabs}
\usepackage{float}
\usepackage{tabularx}

\usepackage{setspace}
\hypersetup{
    colorlinks=true, 
    linkcolor=blue, 
    citecolor=blue, 
    urlcolor=blue,
    pdfauthor={Cheng Zhang} 
}
\usepackage[capitalize]{cleveref}

\journal{Economics Letters}

\begin{document}

\begin{frontmatter}

\title{A Nontrivial Upper Bound on the Out-of-Sample \texorpdfstring{$R^2$}{R2} in Return Forecasting}

\author[inst1,inst2]{Cheng Zhang\corref{cor1}\orcidlink{0000-0002-4150-3371}} 

\ead{zhangcheng01@hbpu.edu.cn, zhang.cheng@northampton.ac.uk}

\cortext[cor1]{Corresponding author}

\affiliation[inst1]{
    organization={Hubei Polytechnic University},
    city={Huangshi},
    postcode={435003}, 
    country={China}
}

\affiliation[inst2]{
    organization={Faculty of Education, Arts, Science and Technology, University of Northampton},
    city={Northampton},
    postcode={NN1 5PH}, 
    country={United Kingdom}
}

\begin{abstract}
This study establishes a nontrivial upper bound on the out-of-sample $R^2$ ($R^2_{\text{OOS}}$) in return forecasting. In particular, we define a coin-flip oracle model that, under the same directional accuracy, theoretically outperforms practical models in terms of MSE. The $R^2_{\text{OOS}}$ of the oracle model, whose analytical expression is a quadratic function of directional accuracy, can therefore serve as a tractable upper bound on the actual $R^2_{\text{OOS}}$. Empirical analyses across multiple forecasting scenarios reveal that the $R^2_{\text{OOS}}$ values of common predictive models are fundamentally bounded by this quadratic function.
\end{abstract}



\end{frontmatter}



\section{Introduction}
\label{sec:intro}

In this study, we aim to explore a nontrivial upper bound on the out-of-sample $R^2$ ($R^2_{\text{OOS}}$) in return forecasting. Prior studies have shown that predictive models typically perform worse than the naive baseline in terms of various error metrics \citep{Meese1983, Kilian2003, Campbell2008, Moosa2013, FotiosPetropoulos2022, Ellwanger2023}, raising the question of whether one can continually improve out-of-sample performance by using more advanced predictive models. Given that the complexity of predictive models contributes little to $R^2_{\text{OOS}}$ values \citep{Welch2008, FotiosPetropoulos2022, Farmer2023}, we argue that a nontrivial upper bound other than $R^2_{\text{OOS}} = 1$ exists.

Since the performance of the unconditional MSE-optimal forecast is intractable, we define a coin-flip oracle model as a proxy for the theoretically best predictive model. In particular, the oracle forecast uses the true conditional expected absolute return at each step, and its predicted sign is generated by a Bernoulli process with a constant probability of sign correctness. Under the same directional accuracy, it theoretically outperforms practical models in terms of MSE. Consequently, the $R^2_{\text{OOS}}$ of this oracle forecast, whose analytical expression is a quadratic function of directional accuracy, provides a tractable upper bound for real-world predictive models. 

By juxtaposing the performance of various predictive models across multiple forecasting scenarios, we observe that the $R^2_{\text{OOS}}$ values of practical models are fundamentally bounded by this quadratic function. The findings of this study also offer a novel perspective on the dependency between conditional mean predictability and sign predictability.
 
\section{Derivation of the Upper Bound}
\label{sec:theory}

\subsection{The Coin-Flip Oracle Model}

Let $r_t = s_t |r_t|$ denote the log return of a financial asset at time $t$, where $s_t \in \{-1, 1\}$ denotes the sign of $r_t$, with zero returns assigned a positive sign. The forecast of a practical model, denoted as $\hat{r}_t^{\text{practical}}$, is given by 
\begin{equation}
\label{eq:r_hat_practical}
\hat{r}_t^{\text{practical}} = \hat{s}_t \hat{m}_t,
\end{equation}
where $\hat{s}_t \in \{-1, 1\}$ denotes the sign of $\hat{r}_t^{\text{practical}}$, and $\hat{m}_t$ denotes the predicted magnitude. Let $\mathbb{I}_t^{\text{practical}}$ denote the indicator of sign correctness for $\hat{s}_t$. Accordingly, the conditional probability of sign correctness, $p_t$, satisfies $\mathbb{P}(\mathbb{I}_t^{\text{practical}} = 1 \mid \Omega_{t-1}) = p_t$, where $\Omega_{t-1}$ denotes the information set available at time $t-1$.

We then define an oracle forecast $\hat{r}_t^{\text{oracle}}$, whose sign forecast $\hat{s}_t^{\text{oracle}}$ is generated by a Bernoulli process with a constant probability $p$ ($p \ge 0.5$) such that $\mathbb{P}(\mathbb{I}_t^{\text{oracle}} = 1 \mid \Omega_{t-1}) = p = \mathbb{E}[p_t]$, where $\mathbb{I}_t^{\text{oracle}}$ is the indicator of sign correctness for $\hat{s}_t^{\text{oracle}}$. The magnitude of $\hat{r}_t^{\text{oracle}}$ under MSE loss is $(2p - 1)\psi_t$, where $\psi_t$ denotes the conditional expected absolute return $\mathbb{E}[|r_t| \mid \Omega_{t-1}]$. Therefore, $\hat{r}_t^{\text{oracle}}$ has the following form:
\begin{equation}
\label{eq:r_hat_oracle}
\hat{r}_t^{\text{oracle}} = \hat{s}_t^{\text{oracle}}(2p - 1)\psi_t.
\end{equation}

Given that $p = \mathbb{E}[p_t]$, we can compare the MSEs of the two types of forecasts under the same directional accuracy. Since $s_t$ and $|r_t|$ are considered conditionally independent given $\Omega_{t-1}$ \citep{Anatolyev2010}, the MSE of $\hat{r}_t^{\text{practical}}$, denoted as $\text{MSE}^{\text{practical}}$, is given by 

\begin{equation}
\label{eq:MSE_practical}
\begin{aligned}
\text{MSE}^{\text{practical}} &= \mathbb{E}[(r_t - \hat{s}_t \hat{m}_t)^2]\\
&= \mathbb{E}[r_t^2] - 2\mathbb{E}\Big[\hat{m}_t \, \mathbb{E}[s_t \hat{s}_t \mid \Omega_{t-1}] \, \mathbb{E}[|r_t| \mid \Omega_{t-1}]\Big] + \mathbb{E}[\hat{m}_t^2]\\
&= \mathbb{E}[r_t^2] - 2\mathbb{E}\big[(2p_t - 1) \psi_t \hat{m}_t\big] + \mathbb{E}[\hat{m}_t^2]\\
&= \mathbb{E}[r_t^2] - 2(2p - 1)\mathbb{E}[\psi_t \hat{m}_t] - 4\operatorname{Cov}(p_t, \psi_t \hat{m}_t) + \mathbb{E}[\hat{m}_t^2].
\end{aligned}
\end{equation}

Moreover, the MSE of $\hat{r}_t^{\text{oracle}}$, denoted as $\text{MSE}^{\text{oracle}}$, is given by
\begin{equation}
\label{eq:MSE_oracle}
\begin{aligned}
\text{MSE}^{\text{oracle}} &= \mathbb{E}[(r_t - \hat{r}_t^{\text{oracle}})^2] \\
&= \mathbb{E}[r_t^2] - 2\mathbb{E}\Big[(2p - 1)\psi_t \, \mathbb{E}\big[s_t \hat{s}_t^{\text{oracle}} |r_t| \bigm| \Omega_{t-1}\big]\Big] + \mathbb{E}\big[(2p - 1)^2 \psi_t^2\big]\\
&= \mathbb{E}[r_t^2] - (2p - 1)^2 \mathbb{E}[\psi_t^2].
\end{aligned}
\end{equation}

Based on \cref{eq:MSE_practical,eq:MSE_oracle}, we can derive the difference between the two MSEs as follows:
\begin{equation}
\label{eq:MSE_difference}
\text{MSE}^{\text{practical}} - \text{MSE}^{\text{oracle}}
= \mathbb{E}\Big[\big(\hat{m}_t - (2p - 1)\psi_t\big)^2\Big] - 4\operatorname{Cov}(p_t, \psi_t \hat{m}_t).
\end{equation}

Since high volatility inflates expected return magnitudes \citep{Merton1980, French1987} while reducing sign predictability \citep{Christoffersen2006}, $p_t$ and $\psi_t \hat{m}_t$ move in opposite directions in response to volatility. Therefore, we have $\operatorname{Cov}(p_t, \psi_t \hat{m}_t) \le 0$, which ensures that $\text{MSE}^{\text{practical}} - \text{MSE}^{\text{oracle}} \ge 0$. Thus, given a directional accuracy $p$, the oracle model theoretically outperforms practical models in terms of MSE.

\subsection{The Out-of-Sample \texorpdfstring{$R^2$}{R2} of the Oracle Model}

According to \cite{Welch2008} and \cite{Gu2020}, as the out-of-sample size approaches infinity (with the zero-return prediction serving as the baseline), the $R^2_{\text{OOS}}$ of $\hat{r}_t^{\text{oracle}}$ can be expressed as follows:
\begin{equation}
\label{eq:R2}
\operatorname{plim} R^2_{\text{OOS}} = 1 - \frac{\mathbb{E}[(r_t - \hat{r}_t^{\text{oracle}})^2]}{\mathbb{E}[(r_t - 0)^2]} = 1 - \frac{\mathbb{E}[(r_t - \hat{r}_t^{\text{oracle}})^2]}{\mathbb{E}[r_t^2]}.
\end{equation}

Since the oracle forecast error is unconditionally orthogonal to the forecast itself, the expected squared realized return can be decomposed into the expected squared forecast and the MSE of $\hat{r}_t^{\text{oracle}}$:
\begin{equation}
\label{eq:variance_decomposition}
\mathbb{E}[r_t^2] = \mathbb{E}[(\hat{r}_t^{\text{oracle}})^2] + \mathbb{E}[(r_t - \hat{r}_t^{\text{oracle}})^2].
\end{equation}

Substituting \cref{eq:variance_decomposition} back into \cref{eq:R2} simplifies $\operatorname{plim} R^2_{\text{OOS}}$ to:
\begin{equation}
\operatorname{plim} R^2_{\text{OOS}} = \frac{\mathbb{E}[(\hat{r}_t^{\text{oracle}})^2]}{\mathbb{E}[r_t^2]}.
\label{eq:R2_ratio}
\end{equation}

We further specify the squared realized return as $r_t^2 = \sigma_t^2 \varepsilon_t$, where $\sigma_t$ is the $\Omega_{t-1}$-measurable conditional volatility and $\varepsilon_t$ is a positive multiplicative error term assumed to be i.i.d.~\citep{Granger1995, Engle2006}. Accordingly, we have $\psi_t = \sigma_t \mathbb{E}[\varepsilon_t^{1/2}]$ and $\psi_t^2 = \sigma_t^2 \big( \mathbb{E}[\varepsilon_t^{1/2}] \big)^2$. Based on \cref{eq:r_hat_oracle}, $\mathbb{E}[(\hat{r}_t^{\text{oracle}})^2]$ can be expressed as follows:
\begin{equation}
\label{eq:numerator}
\begin{aligned}
\mathbb{E}[(\hat{r}_t^{\text{oracle}})^2] &= \mathbb{E}[(2p - 1)^2 \psi_t^2]\\
&= (2p - 1)^2 \mathbb{E}[\sigma_t^2] \, \big( \mathbb{E}[\varepsilon_t^{1/2}] \big)^2.
\end{aligned}
\end{equation}

Using the law of total expectation, we can express $\mathbb{E}[r_t^2]$ as follows:
\begin{equation}
\label{eq:denominator}
\begin{aligned}
\mathbb{E}[r_t^2] &= \mathbb{E}\big[\mathbb{E}[\sigma_t^2 \varepsilon_t \mid \Omega_{t-1}]\big] \\
&= \mathbb{E}[\sigma_t^2] \, \mathbb{E}[\varepsilon_t].
\end{aligned}
\end{equation}

By substituting \cref{eq:numerator,eq:denominator} back into \cref{eq:R2_ratio}, we can analytically express the $R^2_{\text{OOS}}$ of the oracle forecast as a quadratic function of directional accuracy $p$:

\begin{equation}
\label{eq:link}
\operatorname{plim} R^2_{\text{OOS}} = \kappa (2p - 1)^2,
\end{equation}
where $\kappa = \big( \mathbb{E}[\varepsilon_t^{1/2}] \big)^2 \big( \mathbb{E}[\varepsilon_t] \big)^{-1}$.

In the empirical analysis, the $R^2_{\text{OOS}}$ of the oracle forecast can be estimated as follows:
\begin{equation}
\label{eq:benchmark}
R^2_{\text{OOS}} = \hat{\kappa} (2\text{DA} - 1)^2,
\end{equation}
where DA represents the realized out-of-sample directional accuracy, and $\hat{\kappa}$ is the sample estimate of $\kappa$ computed over the out-of-sample period of $T$ steps:
\begin{equation}
\label{eq:kappa_hat}
\hat{\kappa} = \bigg( \frac{1}{T} \sum_{t=1}^T \hat{\varepsilon}_t^{1/2} \bigg)^2 \bigg( \frac{1}{T} \sum_{t=1}^T \hat{\varepsilon}_t \bigg)^{-1}.
\end{equation}

In \cref{eq:kappa_hat}, $\hat{\varepsilon}_t$ is the estimated multiplicative error, given by $\hat{\varepsilon}_t = r_t^2 \hat{\sigma}_t^{-2}$, where $\hat{\sigma}_t$ represents the out-of-sample conditional volatility, which can be estimated using a conditional volatility model such as GARCH(1,1) \citep{Bollerslev2023}.

\section{Empirical Analysis}
\label{sec:Empirical}

\subsection{Data}

With the quadratic function provided by \cref{eq:benchmark} as the upper bound, we now turn to actual financial data to examine whether the performance of practical models is bounded by this theoretical limit. Since sign dynamics are most prevalent at intermediate frequencies \citep{Christoffersen2006}, we retrieve 14 financial time series from Yahoo Finance, each containing weekly closing prices. The details of each time series are provided in \cref{tab:data_summary_stats}. Moreover, each dataset is split into in-sample and out-of-sample sets at varying ratios ranging from 80:20 to 60:40. For each data splitting ratio, one $\hat{\kappa}$ is computed according to \cref{eq:kappa_hat}. The values of $\hat{\kappa}$ for each forecasting scenario are provided in \cref{tab:data_summary_kappa}. We then employ ten conventional predictive models to generate out-of-sample log return forecasts. The model details are provided in \cref{tab:models}. The performance of each model is evaluated using $R^2_{\text{OOS}}$ and DA. In addition, the top 2\% of the absolute log returns in each out-of-sample set are excluded to mitigate the impact of sample noise on performance evaluation \citep{Gu2020}.

\subsection{Results}

\begin{figure}[H]
    \centering
    \includegraphics[width=0.85\textwidth]{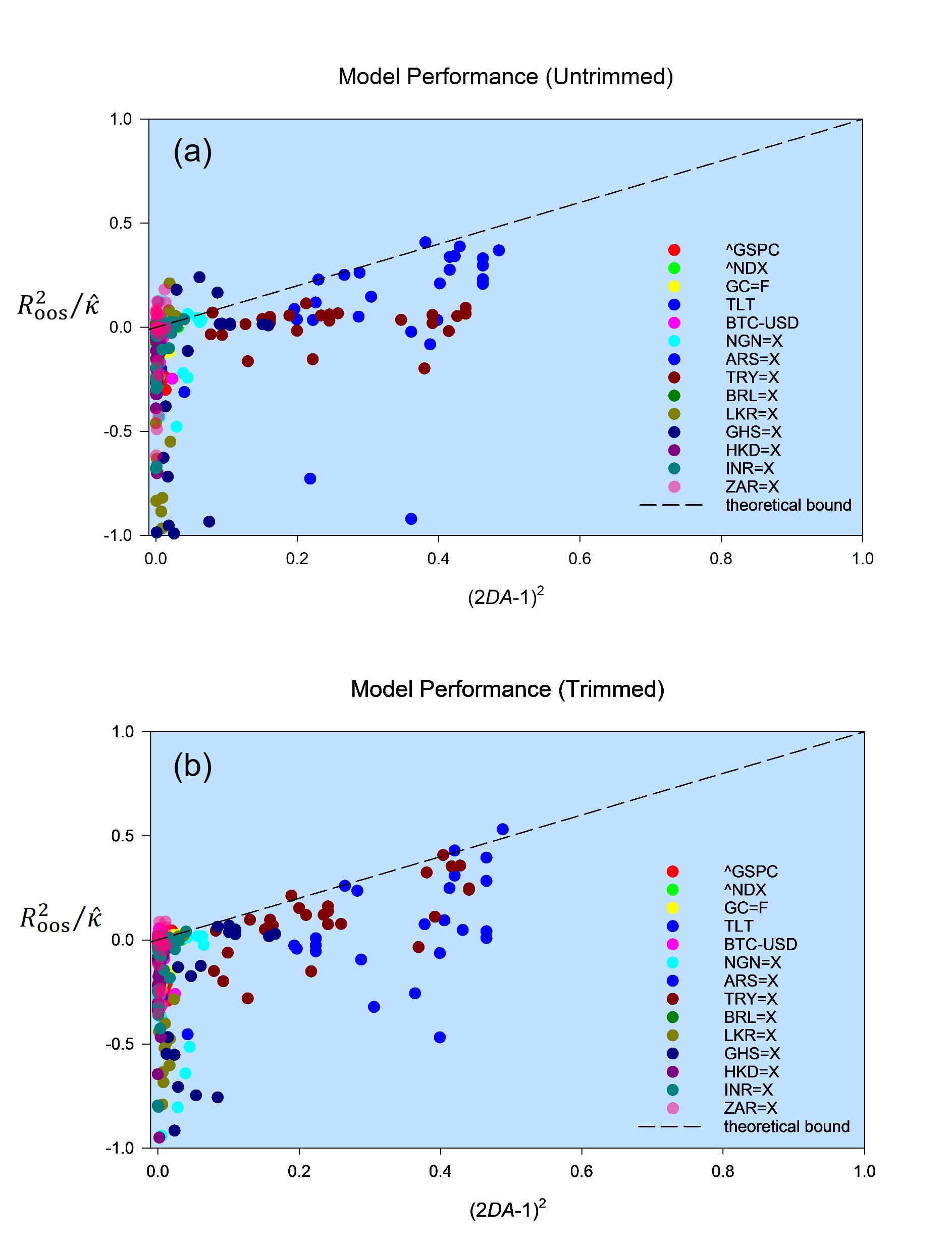}
    \caption{Practical model performance versus the nontrivial upper bound. (a) Untrimmed results. (b) Trimmed results. The dashed line indicates the theoretical upper bound $R^2_{\text{OOS}} = \hat{\kappa} (2\text{DA} - 1)^2$.}
    \label{fig:model_performance_scatter}
\end{figure}

We represent each model's performance as a coordinate pair $\big((2\text{DA} - 1)^2, R^2_{\text{OOS}} / \hat{\kappa}\big)$ and plot all pairs collectively in a single two-dimensional space. This allows the juxtaposition of model performances to cover a wider range of directional accuracies, as shown in \cref{fig:model_performance_scatter}. The reference line represents the nontrivial upper bound given by the oracle model. 

Several observations can be made based on \cref{fig:model_performance_scatter}. First, the data points are fundamentally bounded by the reference line, indicating that model performance is constrained by the quadratic function. Performance falling below the reference line can be attributed to model misspecification or sample variation. Second, many data points have negative y-axis values alongside positive x-axis values, indicating that negative $R^2_{\text{OOS}}$ values are accompanied by modest directional accuracies, which is consistent with the metric disconnect phenomenon reported in empirical studies \citep{Leitch1991, Pesaran1995}. Third, the models can outperform the zero-return baseline when directional accuracy is high. The higher the directional accuracy, the greater the potential $R^2_{\text{OOS}}$ improvement over the naive baseline. Fourth, the results evaluated on the trimmed data show that as sample variation is reduced, spurious deviations above the theoretical bound are largely removed.  

\section{Conclusion}
\label{sec:conclusion}

While $R^2_{\text{OOS}}$ measures the goodness-of-fit of return forecasts, this study shows that it is fundamentally constrained by the nature of the data as well as directional accuracy. Given the quadratic link between $R^2_{\text{OOS}}$ and directional accuracy, minimizing magnitude-based error metrics and maximizing directional accuracy emerge as aligned optimization objectives. Sign predictability does not depend on conditional mean predictability; however, the reverse relationship holds.

\section*{Data Availability}
The data and code are available at 

\url{https://github.com/Zhang-Cheng-76200/R2DA}.

\section*{Acknowledgments}

The author received no specific funding for this research.

\section*{Declaration of interest statement}
The author reports that there are no competing interests to declare.

\section*{Declaration of generative AI and AI-assisted technologies in the manuscript preparation process}

During the preparation of this work, the author used Gemini 3 to improve the readability of the manuscript. After using this tool, the author reviewed and edited the content as needed and takes full responsibility for the content of the published article.

\bibliographystyle{elsarticle-harv}
\bibliography{reference}

\appendix
\section*{Appendix}

\setcounter{table}{0}
\renewcommand{\thetable}{A\arabic{table}}

\begin{table}[H] 
    \centering
    \caption{Data summary and basic statistics}
    \label{tab:data_summary_stats}
    \setlength{\tabcolsep}{4pt} 
    \scriptsize 
    \begin{tabular}{l c c r r r r r}
        \toprule
        Asset & Start Date & End Date & N Obs & Mean & Std Dev & Skewness & Kurtosis \\
        \midrule
        \textasciicircum{}GSPC & 2000-01-08 & 2025-12-27 & 1356 & 0.001149 & 0.024819 & -0.8726 & 7.0927 \\
        \textasciicircum{}NDX  & 2000-01-08 & 2025-12-27 & 1356 & 0.001451 & 0.034445 & -0.7146 & 6.6168 \\
        GC=F    & 2000-09-04 & 2025-12-29 & 1322 & 0.002079 & 0.023565 & -0.2908 & 1.8336 \\
        TLT     & 2002-08-05 & 2025-12-29 & 1222 & 0.000686 & 0.018868 & -0.1596 & 1.3395 \\
        BTC-USD & 2014-09-22 & 2025-12-29 & 589  & 0.009176 & 0.093840 & -0.3472 & 1.9289 \\
        NGN=X   & 2003-12-08 & 2025-12-29 & 1152 & 0.002043 & 0.098612 & 0.2452  & 476.5001 \\
        ARS=X   & 2001-07-16 & 2025-12-29 & 1218 & 0.005978 & 0.044188 & 17.5678 & 393.1939 \\
        TRY=X   & 2005-01-10 & 2025-12-29 & 1095 & 0.003131 & 0.024664 & -1.4312 & 55.4821 \\
        BRL=X   & 2003-12-08 & 2025-12-29 & 1068 & 0.000584 & 0.023694 & -1.9568 & 31.7511 \\
        LKR=X   & 2003-12-08 & 2025-12-29 & 1149 & 0.001011 & 0.014235 & 3.8376  & 64.9548 \\
        GHS=X   & 2007-07-16 & 2025-12-29 & 964  & 0.002574 & 0.068851 & -0.1561 & 298.2477 \\
        HKD=X   & 2001-07-23 & 2025-12-29 & 1242 & -0.000002 & 0.000935 & 0.0129  & 45.0387 \\
        INR=X   & 2003-12-08 & 2025-12-29 & 1149 & 0.000592 & 0.008776 & 0.2098  & 3.0487 \\
        ZAR=X   & 2003-12-08 & 2025-12-29 & 1152 & 0.000841 & 0.029972 & -0.6950 & 24.3722 \\
        \bottomrule
    \end{tabular}
\end{table}

\vspace{1em} 

\begin{table}[H] 
    \centering
    \caption{Raw and trimmed $\hat{\kappa}$ across different training split ratios}
    \label{tab:data_summary_kappa}
    \scriptsize 
    \begin{tabular}{l rr rr rr}
        \toprule
        & \multicolumn{2}{c}{80\% Split} & \multicolumn{2}{c}{70\% Split} & \multicolumn{2}{c}{60\% Split} \\
        \cmidrule(lr){2-3} \cmidrule(lr){4-5} \cmidrule(lr){6-7}
        Asset & Raw $\hat{\kappa}$ & Trimmed $\hat{\kappa}$ & Raw $\hat{\kappa}$ & Trimmed $\hat{\kappa}$ & Raw $\hat{\kappa}$ & Trimmed $\hat{\kappa}$ \\
        \midrule
        \textasciicircum{}GSPC & 0.6051 & 0.6388 & 0.5690 & 0.6296 & 0.5629 & 0.6111 \\
        \textasciicircum{}NDX  & 0.6304 & 0.6642 & 0.5994 & 0.6498 & 0.5937 & 0.6431 \\
        GC=F    & 0.5944 & 0.6225 & 0.5728 & 0.5968 & 0.5859 & 0.6158 \\
        TLT     & 0.6271 & 0.6540 & 0.6017 & 0.6328 & 0.5968 & 0.6340 \\
        BTC-USD & 0.5634 & 0.5918 & 0.4860 & 0.5464 & 0.5164 & 0.5693 \\
        NGN=X   & 0.1017 & 0.4140 & 0.1330 & 0.3125 & 0.1225 & 0.2525 \\
        ARS=X   & 0.0786 & 0.4805 & 0.0933 & 0.4910 & 0.1209 & 0.4460 \\
        TRY=X   & 0.3387 & 0.3701 & 0.4022 & 0.4398 & 0.4132 & 0.4996 \\
        BRL=X   & 0.6205 & 0.6545 & 0.6117 & 0.6456 & 0.6075 & 0.6406 \\
        LKR=X   & 0.0944 & 0.4206 & 0.1154 & 0.4600 & 0.1499 & 0.4564 \\
        GHS=X   & 0.2163 & 0.3899 & 0.1854 & 0.3543 & 0.1988 & 0.3810 \\
        HKD=X   & 0.4880 & 0.5414 & 0.4575 & 0.5054 & 0.4087 & 0.5261 \\
        INR=X   & 0.5086 & 0.5385 & 0.5013 & 0.5500 & 0.5284 & 0.5653 \\
        ZAR=X   & 0.2534 & 0.4830 & 0.3021 & 0.5150 & 0.3440 & 0.5743 \\
        \bottomrule
    \end{tabular}
\end{table}

\vspace{1em}

\begin{table}[H] 
    \centering
    \caption{Configurations for the predictive models}
    \label{tab:models}
    \scriptsize 
    \begin{tabular}{l p{0.7\textwidth}} 
        \toprule     
        \textbf{Model} & \textbf{Description \& Hyperparameter Setup} \\
        \midrule

        \textbf{Mean} & 
        Rolling historical mean using an 8-period window. \\
        \addlinespace

        \textbf{AutoARIMA} & 
        Nonseasonal, stationary ARIMA automatically selected via stepwise search. \\
        \addlinespace

        \textbf{AR-GARCH} & 
        AR(1) conditional mean with a GARCH(1,1) conditional variance equation. \\
        \addlinespace

        \textbf{Ridge} & 
        8-period lag, standardized features. L2 penalty $\alpha = 50.0$. \\
        \addlinespace

        \textbf{ElasticNet} & 
        8-period lag, standardized features. Penalty $\alpha = 0.01$, L1 ratio = 0.5. \\
        \addlinespace

        \textbf{SVR} & 
        8-period lag, standardized features. RBF kernel, $C = 0.1$, and $\epsilon = 0.01$. \\
        \addlinespace

        \textbf{RF} & 
        8-period lag. 100 trees, max depth = 3, min samples per leaf = 10. \\
        \addlinespace

        \textbf{XGB} & 
        8-period lag. 50 trees, max depth = 2, learning rate = 0.05, L2 $\lambda = 10.0$, L1 $\alpha = 5.0$. \\
        \addlinespace

        \textbf{MLP} & 
        8-period lag, standardized data. 1 hidden layer (4 nodes), $\tanh$ activation, L2 $\alpha = 0.1$, Adam optimizer, early stopping. \\
        \addlinespace

        \textbf{RNN} & 
        8-period lag, standardized data. 1 recurrent layer (4 units), $\tanh$ activation, L2 penalty = 0.05, dropout = 0.2, early stopping. \\
        \bottomrule
    \end{tabular}
\end{table}

\end{document}